\documentclass[aps, groupedaddress, superscriptaddress, prl, reprint]{revtex4-1}
\usepackage{graphicx}
\usepackage{setspace,transparent}
\usepackage{color,float,calligra}
\usepackage{fontenc,subcaption,ragged2e,amsmath,stackengine}
\usepackage{mathtools,amssymb,amssymb,nicefrac,tikz}
\usepackage{soul,mathrsfs}
\usepackage[font=small,labelfont=bf]{caption}
\usepackage[utf8]{inputenc}

\usepackage{etoolbox}
\patchcmd{\thebibliography}{\chapter*}{\section*}{}{}

\DeclareCaptionJustification{justified}{\justifying}
\DeclareMathAlphabet{\mathpzc}{OT1}{pzc}{m}{it}
\DeclareMathAlphabet{\mathcalligra}{T1}{calligra}{m}{n}

\DeclareCaptionJustification{justified}{\justifying}
\begin{document}

\title{Critical stretching of mean-field regimes in spatial networks}
\author{Ivan Bonamassa}
\affiliation{Department of Physics, Bar Ilan University, Ramat Gan, Israel}
\author{Bnaya Gross}
\affiliation{Department of Physics, Bar Ilan University, Ramat Gan, Israel}
\author{Michael M. Danziger}
\affiliation{Network Science Institute, Northeastern University, Boston, USA}
\author{Shlomo Havlin}
\affiliation{Department of Physics, Bar Ilan University, Ramat Gan, Israel}
\date{\today}

\begin{abstract}
We study a spatial network model with exponentially distributed link-lengths on an underlying grid of points, undergoing a structural crossover from a random, Erd\H{o}s--R\'enyi graph to a $2D$ lattice at the characteristic interaction range $\zeta$.
We find that, whilst far from the percolation threshold the random part of the incipient cluster scales linearly with $\zeta$, close to criticality it extends in space until the universal length scale $\zeta^{3/2}$ before crossing over to the spatial one. 
We demonstrate this {\em critical stretching} phenomenon in percolation and in dynamical processes, and we discuss its implications to real-world phenomena, such as neural activation, traffic flows or epidemic spreading.
\end{abstract}

\maketitle
Two decades after its inception, Network Theory has grown to become an exhaustive framework for taming the complexity of a multitude of systems~\cite{books}, enabling researchers to quantitatively predict and to optimally design their collective organization at different scales~\cite{NetBioMed,community,multi,Yan017}. 
Despite the advances made in the study of critical phenomena~\cite{Collective} on random networks, their analysis in the presence of a geometric embedding has remained an important theoretical challenge, so far grounded on few analytical results~\cite{Bun008,bra010,bar011,Li2011,Bart18}.
This represents today a fundamental barrier in reaching a thorough understanding of many complex systems of great interest, like networks of neurons~\cite{Betzel2017,breakspear17}, communication~\cite{Lam008} and transportation systems~\cite{Li2018}, or power-grids~\cite{Mot017}, whose functioning is strongly intertwined with their spatial constraints. \\
\indent 
This Letter aims to explore the effects that a tunable characteristic link length $\zeta$ has on the universality of critical phenomena in a realistic model of a spatial network.
Physical situations of experimental relevance, in fact, often exhibit at least two relevant scales: one is determined by the correlation length $\xi$, and it is therefore related to the experimental conditions (i.e.\,the control parameters of the system), while a second one, say $\xi_*$, is defined by the interaction length (related to the characteristic scale $\zeta$), and it is instead intrinsic to the system.
In cases where these two scales are comparable, one does not observe the expected critical behavior, but rather a {\em crossover}~\cite{Bin001,Pel002} from a high-dimensional, mean-field behavior (small $\xi/\xi_*$ or, equivalently, far from criticality) to a low-dimensional one (large $\xi/\xi_*$, or close to criticality).\\
\indent 
To cope with these realistic scenarios, we consider here a variant on a grid of the so-called Waxman model~\cite{Wax988}, where $N=L^2$ nodes are regularly distributed on a square of linear size $L$ with spacing $a\equiv1$, and edges are drawn with probabilities $\mathpzc{P}_{ij}\propto \kappa\,e^{-d_{ij}/\zeta}$, for every pair of nodes $i,\,j$ placed at the Euclidean distance $d_{ij}$. 
Like in Erd\H{o}s-R\'enyi (ER) networks, the parameter $\kappa\equiv\langle k\rangle/(N-1)$ controls the total density of links, with $\langle k\rangle$ the average connectivity, while $\zeta\in(0,L]$ is a characteristic link-length. 
Although simplistic, the exponential wiring cost function of this model correctly fits observed communication systems~\cite{Wire009}, like the Internet~\cite{Zeg996,Yook2002} or transport networks~\cite{Hal014,Dan016}, interactions based on similarity in social networks~\cite{Watts2002}, and recent experimental results~\cite{Brain1,Bull012,Brain2,Brain3} support the existence of scale-invariant organizational principles in the mammalian brain based on it.\\
\indent 
This disordered lattice is a natural benchmark for investigating crossover phenomena in spatial networks.
Depending on the values of $\zeta$, in fact, our model smoothly crosses over between two regimes: for $\zeta\sim\mathcal{O}(L)$, all link lengths are equally probable (i.e.\,$\mathpzc{P}_{ij}\sim \kappa$) and we recover a random, ER-like structure, whilst for $\zeta\sim\mathcal{O}(1/\sqrt{\rho})$ (where $\rho$ is the average density of nodes), long links are prohibited, leading instead to a lattice-like topology. 
By performing site percolation~\cite{Con982,Stauf14,BuH012}, we disclose an intriguing scenario: while far from criticality the network undergoes a structural crossover from a small- to a large-world at the characteristic scale $\zeta$, close to the percolation threshold the random part of the incipient cluster extends in space until the intrinsic scale $\xi_*=\zeta^{3/2}$. 
We call this phenomenon {\em critical stretching}, and we explore its properties by means of scaling arguments and simulations.
In particular, because the structure affects the collective phenomena occurring on it, the critical stretching should be reflected in processes evolving on this spatial network. 
We verify this claim on an SIR (susceptible-infective-removed) process~\cite{epidemics}, where the mean-field (diffusive) regime of the outbreak front's velocity undergoes a temporal critical stretching, making the epidemic invade the system at ``slow'' rates for longer times.\vspace{-0.1cm}

\begin{figure}[bt]
	\centering	
	\begin{tikzpicture}[      
	every node/.style={anchor=south west,inner sep=0pt},
	x=1mm, y=1mm,]   
	\node (fig1) at (0,-20)
	{\includegraphics[scale=0.42]{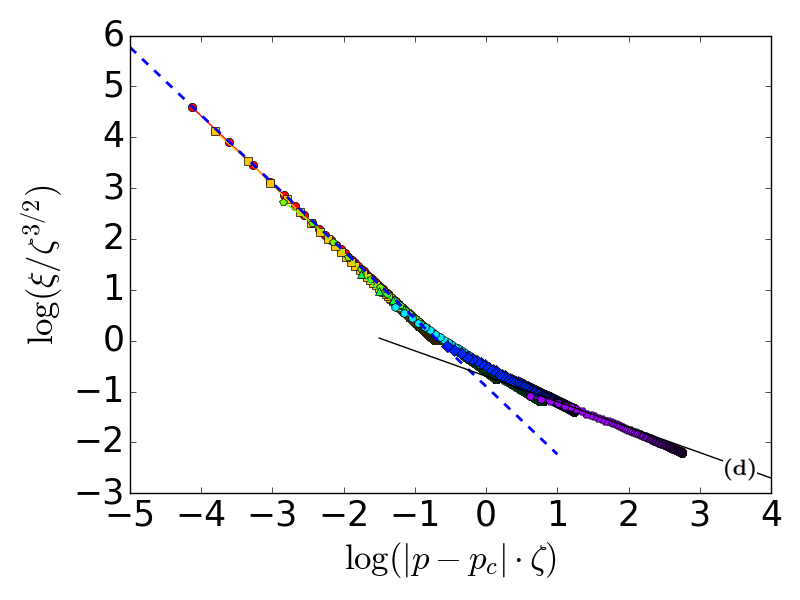}};
	\node (fig2) at (+15,-8.0)
	{\includegraphics[scale=0.23]{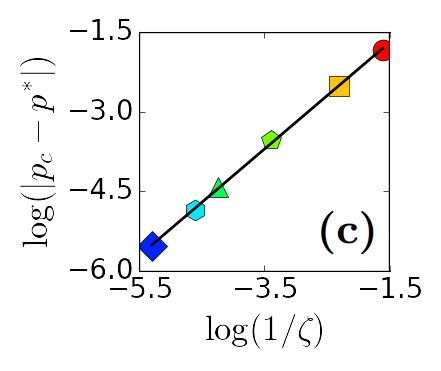}};
	\node (fig3) at (41,8)
	{\includegraphics[scale=0.205]{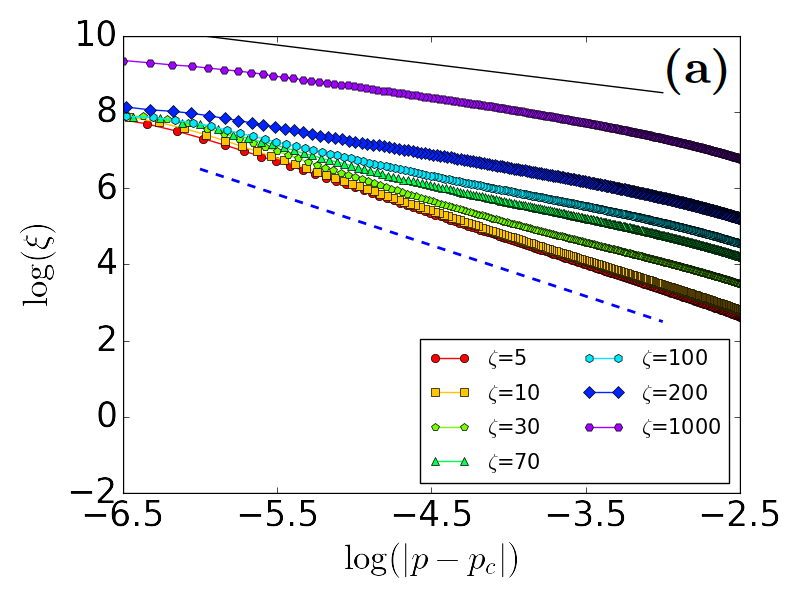}};
	\node (fig4) at (+48,+14)
	{\includegraphics[scale=0.092]{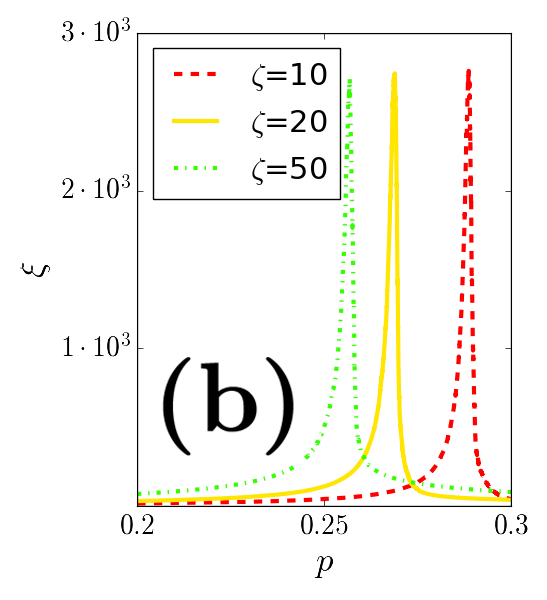}};
	\end{tikzpicture}\vspace*{-0.35cm}
	\caption{\textbf{Crossover of the correlation length.} 
	\textbf{(a)} Below criticality $\xi\!\sim\!(p_c-p)^{-\nu}$, where the lattice exponent $\nu_{2D}=4/3$ (dashed line) appears at large distances (i.e.\,$|p-p_c|\ll1$), while the ER one $\nu_{MF}=1/2$ (thick line) is found otherwise. 
	For $\zeta$ large enough, the scaling of $\xi$ crosses over between these extreme cases. 
	\textbf{(b)} Shift of $p_c$ for increasing $\zeta$; notice that $p_c\simeq0.25$ for $\zeta\gg1$. 
\textbf{(c)} Collapse of the data in \textbf{(a)} according to Eq.~\eqref{eq:1}. 
\textbf{(d)} The fitted slope yields $\mathpzc{p}_*=\zeta^{-\omega}$ with $\omega\simeq1.01$.
	The curves are obtained for $N=10^{8}$, averaging over $200$ realizations. In this and all following figures, the $\zeta$ values are those specified in the legend, unless otherwise stated.}\vspace*{-0.7cm}
	\label{fig:1}
\end{figure}

\paragraph*{\underline{Crossover and critical stretching.}} 
To investigate the structural crossover, we perform site percolation by randomly removing a fraction $1-p$ 
of the network's nodes, for given average degree $\langle k\rangle=4$ and varying characteristic length $\zeta$.
We focus on values of $\zeta\in[1,L/2)$, where the upper bound avoids the re-drawing of edges and the lower one is set by the lattice-dominated regime~\cite{note1}.
For increasing $\zeta$, a rapid monotonic decrease of the percolation threshold from the lattice value $p_c^{2D}\simeq0.59$ to the ER one $p_c^{MF}=1/\langle k\rangle$ is observed (Fig.~\ref{fig:1}b, and Ref.~\cite{Dan016}), signaling a structural crossover between the two topologies. 
To analyze the response the scaling of percolation observables have to this variation in $p$, we begin by considering the correlation length $\xi^2(p):=\int r^2 g(r,p)\mathrm{d}r/\int g(r,p)\mathrm{d}r$, where $g(r,p)$ is the pair connectivity function for a given occupation probability $p$.
By definition, $\xi$ measures the mean distance between sites belonging to the same cluster, and it is expected to diverge as $p$ approaches $p_c$ like $\xi\sim|p-p_c|^{-\nu}$. 
Numerical results reported in log-log scale in Fig.~\ref{fig:1}a display excellent agreement with the critical exponents expected in the two limiting cases, i.e.\,$\nu_{2D}=4/3$ for $\zeta\sim\mathcal{O}(1)$ and $\nu_{MF}=1/2$ for $\zeta\sim\mathcal{O}(L)$, further showing that a crossover from the former to the latter regime occurs closer and closer to $p_c$ for increasing $\zeta$.\\
\indent 
Such critical crossover can be understood by invoking the Levanyuk-Ginzburg (LG) criterion~\cite{LGcrit}.
Since the spectrum of the spatial network model here considered satisfies the classical LG assumptions~\cite{bra010}, we expect critical fluctuations to be negligible as long as $\mathpzc{p}^{3-d/2}\zeta^d\gg1$, where $\mathpzc{p}\equiv |p/p_c-1|$ is the displacement of $p$ from its critical value. 
To obtain this condition, we noticed that the distance up to which the mean-field regime extends in the incipient cluster is $\xi_{MF}\propto\zeta{\xi_{\ell}}^{1/2}$, where $\xi_\ell(\mathpzc{p})\sim\mathpzc{p}^{-1}$ is the (dimensionless) correlation length in chemical space, and the factor $\zeta$ takes into account the characteristic length of bonds~\cite{note2}. 
In $d=2$ the crossover should then take place at the probability $\mathpzc{p}_*\equiv\zeta^{-1}$ and, in light of the above, at the intrinsic scale $\xi_*\equiv\zeta^{3/2}$.
For such variation to occur, the amplitude modulating the scaling of $\xi$ with $\mathpzc{p}$ must itself become anomalous in $\zeta$ as soon as critical fluctuations set in, so that $\xi_{2D}\!\propto\!\zeta^{\psi}\mathpzc{p}^{-4/3}$. 
The exponent $\psi$ can then be found by invoking a smooth variation at $p_*$, i.e.\,$\xi_{MF}|_{p_*}=\xi_{2D}|_{p_*}$, yielding $\psi=1/6$. 
Factoring out $\xi_*$, as well as $\mathpzc{p}/\mathpzc{p}_*$, we obtain the generalized scaling
\begin{equation}\label{eq:1}
\frac{\xi(\mathpzc{p},\zeta)}{\xi_*(\zeta)}=\left(\frac{\mathpzc{p}}{\mathpzc{p}_*(\zeta)}\right)^{\!\!-\nicefrac{1}{2}}\!\!\mathcal{F}\left(\frac{\mathpzc{p}}{\mathpzc{p}_*(\zeta)}\right),
\end{equation}
where $\mathcal{F}(\mathpzc{x})\propto \mathpzc{x}^{\-5/6}$ for $\mathpzc{x}\ll1$, and constant otherwise. 
With these predictions in hand, the data presented in Fig.~\ref{fig:1}a have been rescaled and restricted to those parts not plagued by finite-size effects or lying outside the critical window. 
The data collapse in Fig.~\ref{fig:1}c,d supports our scaling arguments and highlights an interesting prediction of the LG criterion. 
Although one would have naively expected the random to spatial crossover to occur at the characteristic scale $\zeta$---as it does far from criticality---here it stretches until $\zeta^{3/2}$. 
This scale can be understood as the network analogue at the crossover of what Binder calls ``thermodynamic length'' 
in spin systems~\cite{Bin001}.
However, in contrast to thermal phase transitions, the scale $\xi_*$ identifies in our problem a ``structural length'' at which geometric fluctuations set in.
This means that $\xi_*$ does not only rule the critical scaling of the percolation observables, but also the network's geometry at the onset of the giant component, stretching with $\zeta$ its high-dimensional structure. 
For this reason, we call this phenomenon {\em critical stretching} and in what follows we characterize its geometric and dynamic properties.

\begin{figure}[bt]
	\centering
	\begin{tikzpicture}[      
	every node/.style={anchor=north west,inner sep=0pt},
	x=1mm, y=1mm,]   
	\node (fig1) at (0,0)
	{\hspace*{+0.05cm}\includegraphics[scale=0.255]{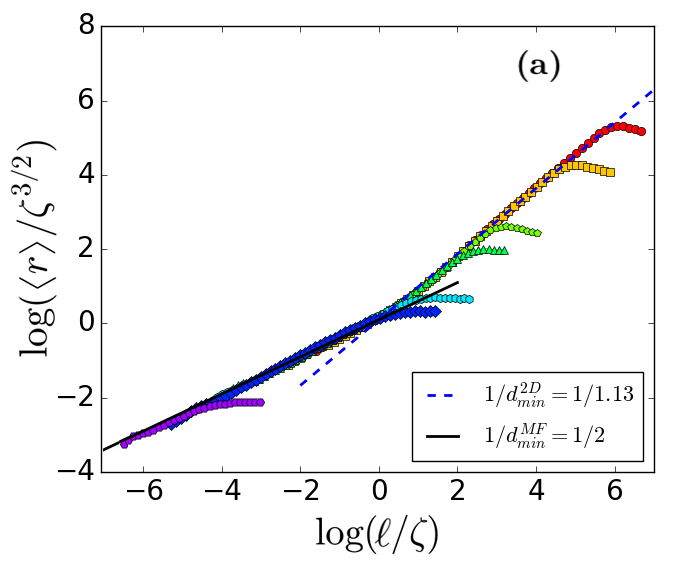}};
	\node (fig2) at (+8.0,-3.5)
	{\includegraphics[scale=0.13]{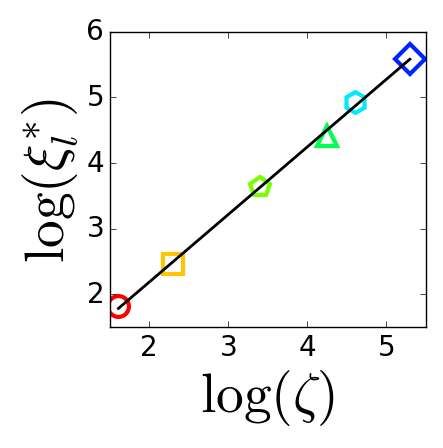}};
	\end{tikzpicture}\hspace*{-0.12cm}
	\begin{tikzpicture}[      
	every node/.style={anchor=south east,inner sep=0pt},
	x=1mm, y=1mm,]   
	\node (fig1) at (0,0)
	{\includegraphics[scale=0.263]{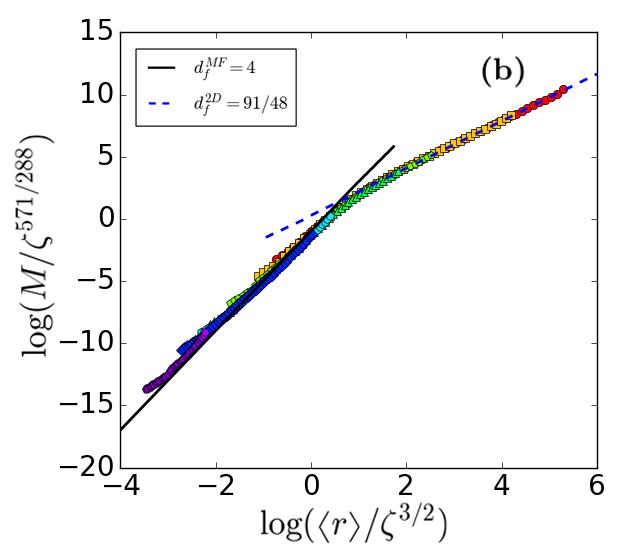}};
	\node (fig2) at (-3.8,6.5)
	{\includegraphics[scale=0.125]{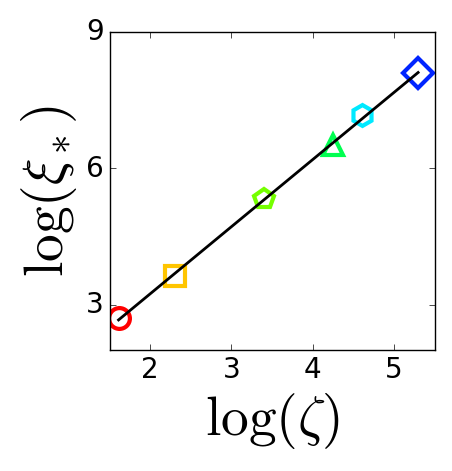}};
	\end{tikzpicture}\vspace*{-0.07cm}\\
	\hspace{+0.05cm}\begin{tikzpicture}[      
	every node/.style={anchor=south east,inner sep=0pt},
	x=1mm, y=1mm,]   
	\node (fig1) at (0,0)
	{\includegraphics[scale=0.285]{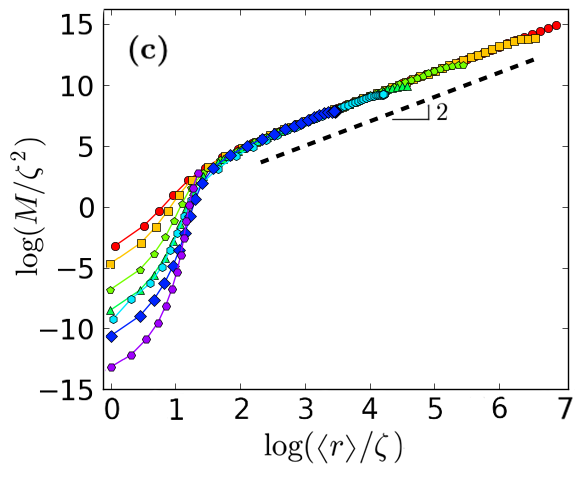}};
	\node (fig2) at (-2.5,+7.5)
	{\includegraphics[scale=0.13]{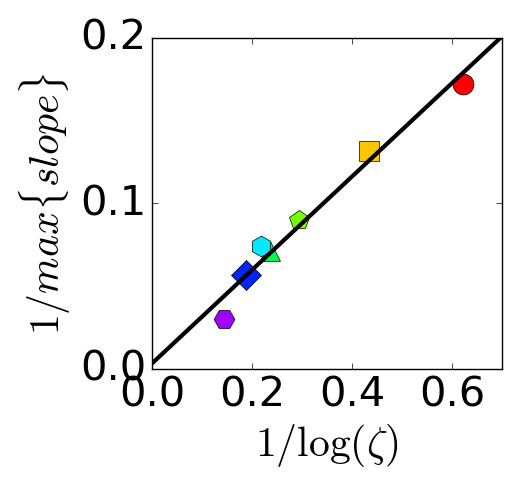}};
	\end{tikzpicture}
	\begin{tikzpicture}[      
	every node/.style={anchor=south east,inner sep=0pt},
	x=1mm, y=1mm,]   
	\node (fig1) at (0,0)
	{\includegraphics[scale=0.315]{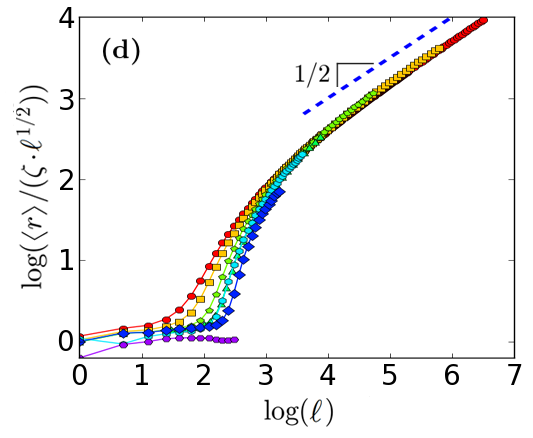}};
	\node (fig2) at (-3.5,+7.5)
	{\includegraphics[scale=0.11]{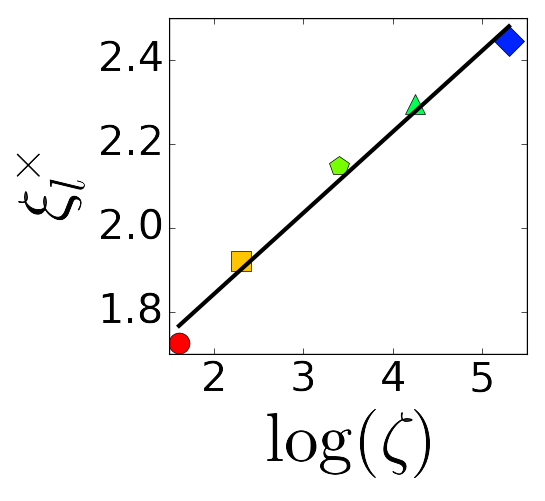}};
	\end{tikzpicture}\vspace*{-0.25cm}
	\caption{\textbf{\small{Geometric crossover and stretching}.}
	\textbf{(a)} Scaled data of $\langle r\rangle(\ell)$ at $p_c$ as predicted by Eq.~\eqref{eq:3}; a linear fit (inset) yields $\xi_\ell^*=\zeta^{\psi}$, $\psi\simeq1.03$.
	\textbf{(b)} Mass scaling at $p_c$: scaled data are consistent with Eq.~\eqref{eq:4}. 
	(Inset) The fitted slope of $\xi_*=\zeta^{\phi}$ yields $\phi\simeq1.48$.
	\textbf{(c)} Mass scaling at $p=1$: for $r\geq\zeta$ the network has the lattice's geometry, while for $r\leq\zeta$ it embeds onto a Euclidean space whose dimension increases with $\zeta$. 
	(Inset) Inverse maximal-slope of the mass scaling in the small-world regime vs.\,$1/\log(\zeta)$: the network's dimension grows unbounded for $\zeta\gg1$.
	\textbf{(d)} Shortest-path scaling at $p=1$. 
	(Inset) Crossover value $\xi_\ell^{\times}\simeq\log\zeta$ in semi-log scale.
	Results are for $N=10^8$, averaged over $200$ realizations.}\vspace{-0.5cm}
	\label{fig:3}
\end{figure}

\paragraph*{\underline{Geometric stretching at criticality.}} 
From a geometric standpoint, the crossover scaling of the correlation length given in Eq.~\eqref{eq:1} implies a non-trivial structural variation of the large-scale organization of the network close to its percolation threshold. 
To put this claim on a firmer basis, we consider the scaling of the pair connectivity function $g(r,\mathpzc{p},\zeta)\propto r^{-d+2-\eta}e^{-r/\xi(\mathpzc{p},\zeta)}$, where $\eta$ is the anomalous dimension. 
The correlation length $\xi(\mathpzc{p},\zeta)$ is itself the representative of a crossover in the incipient cluster, separating a self-similar organization defined by the power-law decay $g(r)\sim r^{-d+2-\eta}$, from a ``non-critical'' one marked instead by a decreasing exponential~\cite{Stauf14}. 
Then again, we know from Eq.~\eqref{eq:1} that $\xi(\mathpzc{p},\zeta)$ crosses over between the mean-field and the lattice singular behaviors, depending on how close one is to the percolation threshold. 
Let us then fix the relative occupation probability to a value $\mathpzc{p}\ll\zeta^{-1}$, so that geometric fluctuations are free to set in above the intrinsic scale $\xi_*=\zeta^{3/2}$. 
In turn, this means that $\xi_\mathpzc{p}\gg\zeta^{3/2}$ and so the cutoff $e^{-r/\xi(\mathpzc{p},\zeta)}$ becomes sig-\vspace{-0.05cm}
\noindent 
nificant only if $r\sim\mathcal{O}(\zeta^{3/2})$. 
Therefore, reading $g_\mathpzc{p}(r,\zeta)$ as a function of $r$ for fixed $\mathpzc{p}\ll\zeta^{-1}$ and variable values of $\zeta$, we advance the crossover scaling ansatz
\begin{equation}\label{eq:2}
\frac{g_\mathpzc{p}\big(r,\zeta\big)}{g_*(\zeta)}=\left(\frac{r}{\xi_*(\zeta)}\right)^{\!-4}\mathcal{G}\bigg(\frac{r}{\xi_*(\zeta)},\frac{\xi_\mathpzc{p}}{\xi_*(\zeta)}\bigg)
\end{equation}
\noindent 
where $g_*(\zeta)\equiv \zeta^{-6}$, and $\mathcal{G}(\mathpzc{x},\mathpzc{y})\propto \mathpzc{x}^{91/24}e^{-\mathpzc{x}/\mathpzc{y}}$ for $\mathpzc{x}\gg1$ and constant otherwise, with $\mathpzc{y}\gg1$. 
To obtain Eq.~\eqref{eq:2}, we have rescaled the (dimensionless) argument of the exponential cutoff in order to take into account explicitly the ratio $r/\xi_*$, by choosing as a reference the mean-field critical behavior, namely $g_{MF}(r)/g_*\sim(r/\xi_*)^{-4}$.\\
\indent 
Eq.~\eqref{eq:2} states that geometric fluctuations in our model are negligible as long as $r\ll\zeta^{3/2}$. 
The intriguing aspect of this geometric counterpart of the LG criterion relies on the fact that it does not depend on the probability $\mathpzc{p}$, i.e.\,it is consistent {\em at} criticality.
As a consequence, $\xi_*$ identifies the intrinsic length at which the geometry of the giant critical cluster crosses over from a high-dimensional fractal to a low-dimensional one. 
To investigate this {\em geometric crossover}, we focus on two exponents: \emph{1)} the shortest-path fractal dimension $d_{min}$, 
and \emph{2)} the fractal dimension $d_f$ of the incipient cluster~\cite{BuH012}. 
An important byproduct is that the critical stretching phenomenon can be then tested by comparing the results obtained from the above measurements {\em at} and {\em away from} criticality.
Let us start from the behavior at $p=p_c$. 
\emph{1)} By definition, $d_{min}$ regulates the scaling of the average Euclidean distance $\langle r\rangle$ between sites in the giant cluster separated by the chemical distance 
$\ell$, so that $\ell\sim\langle r\rangle^{d_{min}}$ for $\langle r\rangle$ large enough. 
For the sake of clarity, let us consider here the reciprocal scaling $\langle r\rangle\sim\ell^{1/d_{min}}$, for $\ell$ large enough. 
In chemical space, the length $\xi_*$ translates~\cite{note2} in the quantity $\xi^*_\ell\equiv\lfloor\zeta\rfloor$ (where the brackets denote the floor function) which, consistently with the scaling ansatz in Eq.~\eqref{eq:2}, identifies the intrinsic chemical length at which geometric fluctuations induce a structural crossover.
In defining $d_{min}$, we must thus consider two regimes: $\ell\ll\xi_\ell^*$ where $d_{min}^{MF}=2$ (i.e.\,random walk), and $\ell\gg\xi_\ell^*$ where instead $d_{min}^{2D}\simeq1.13077(2)$~\cite{dmin}. 
Merging the two regimes into a single equation, we propose the scaling
\begin{equation}\label{eq:3}
\frac{\langle r\rangle(\ell,\zeta)}{\xi_*(\zeta)}=\bigg(\frac{\ell}{\xi_\ell^*(\zeta)}\bigg)^{\!\!\nicefrac{1}{2}}\mathcal{R}\bigg(\frac{\ell}{\xi_\ell^*(\zeta)}\bigg),
\end{equation}
\noindent 
where $\mathcal{R}(\mathpzc{x})\propto\mathpzc{x}^{1/{d_{min}^{2D}}-1/2}$ for $\mathpzc{x}\gg1$ and constant otherwise. 
We tested Eq.~\eqref{eq:3} by using a breadth-first search algorithm whose results, displayed in Fig.~\ref{fig:3}(a), support our scaling arguments. 
\emph{2)} To investigate the mass scaling $M\sim r^{d_f}$ at $p_c$, we first need to determine the dependence on $\zeta$ of the incipient cluster's amplitude. 
This quantity is obtained by calculating the $\zeta$-correction to the size $S_\infty$ of the giant cluster in the spatial regime which, in line with the LG criterion, scales as $S_\infty^{2D}\propto\mathpzc{B}(\zeta)\mathpzc{p}^{5/36}$, where $\mathpzc{B}(\zeta)\equiv\zeta^{-31/36}$. 
Knowing $\mathpzc{B}(\zeta)$, we can find the $\zeta$-correction to the mass scaling in the spatial regime, i.e.\,$M_{2D}\propto \mathpzc{B}(\zeta)r^{91/48}$ for $r\gg\xi_*$. 
Merging this behavior with the one expected in the high-dimensional one, i.e.\,$M_{MF}\sim r^{4}$~\cite{BuH012}, we write the ansatz
\begin{equation}\label{eq:4}
\frac{M(r,\zeta)}{M_*(\zeta)}=\bigg(\frac{r}{\xi_*(\zeta)}\bigg)^{\!4}\mathcal{M}\bigg(\frac{r}{\xi_*(\zeta)}\bigg),
\end{equation}
\noindent 
where $\mathcal{M}(\mathpzc{x})\propto\mathpzc{x}^{-101/48}$ for $\mathpzc{x}\gg1$, constant otherwise, and $M_*(\zeta)\equiv\zeta^{571/288}$ is the giant cluster's mass at $\xi_*$. 
Eq.~\eqref{eq:4} states that the fractal geometry of the critical cluster in our network crosses over from the one of an ER graph~\cite{CoH004} (whose fractal dimension is $d_f^{MF}=4$) to\vspace*{-0.06cm} 
\noindent 
that of a square lattice (for which $d_f^{2D}=91/48$), where\vspace*{-0.06cm} 
\noindent 
the former stretches up until the intrinsic length $\zeta^{3/2}$. 
As shown in Fig.~\ref{fig:3}(b), Eq.~\eqref{eq:4} exquisitely matches the numerical data, which we obtained by means of the spatial cluster growing algorithm introduced in Ref.~\cite{Li2011}.\\
\indent 
To complete the picture, we must verify that {\em far} from the percolation threshold, $\xi_\times\equiv\zeta$ is indeed the characteristic length at which the network's structure crosses over from an ER graph to a $2D$ lattice. 
Let us start from the scaling \emph{2)} of the mass at $p=1$. 
In this case, the Euclidean dimension $d=2$ is expected to regulate the scaling of $M$ for distances $r\sim\mathcal{O}(\xi_\times)$, while an infinite-dimensional (small-world) behavior should be observed for $r\sim\mathpzc{o}(\xi_\times)$. 
The numerical results in Fig.~\ref{fig:3}(c) corroborate this intuition, further showing (Fig.~\ref{fig:3}(c), inset) that below $\xi_\times$ the random structure embeds onto an Euclidean space whose dimension increases unbounded with $\zeta$. 
The measurement \emph{1)} of the shortest-path's scaling at $p=1$ supports this large- to small-world crossover.
In fact, for $r\lesssim\xi_\times$ \emph{any} path shall behave like a random walk~\cite{BuH012}, i.e.\,$\langle r\rangle\propto\zeta\ell^{1/2}$, where the prefactor takes into account the average link length in this regime. 
\begin{figure}[bt]
	\centering
	\begin{tikzpicture}[      
	every node/.style={anchor=south east,inner sep=0pt},
	x=1mm, y=1mm,]   
	\node (fig1) at (0,0)
	{\includegraphics[scale=0.365]{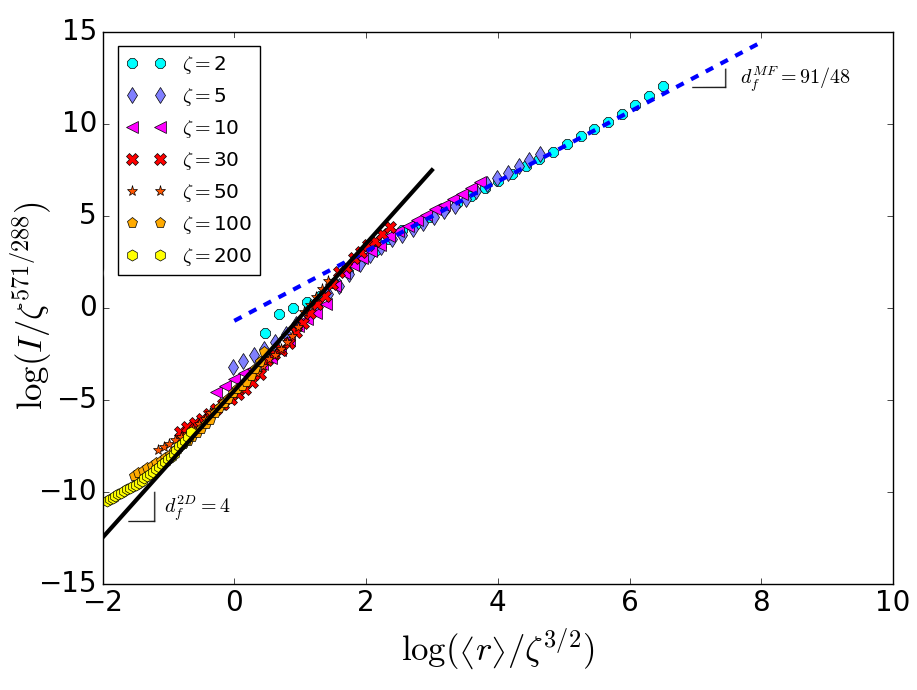}};
	\node (fig2) at (-3.0,11.0)
	{\includegraphics[scale=0.310]{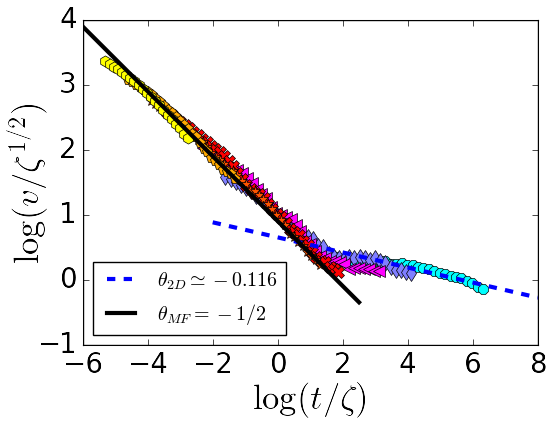}};
	\node (fig3) at (-6,+25)
	{\includegraphics[scale=0.145]{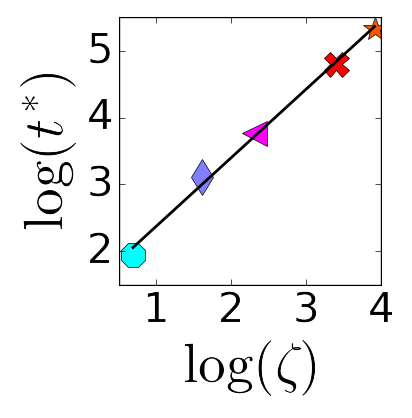}};
	\end{tikzpicture}\vspace*{-0.35cm}
	\caption{\textbf{Dynamical stretching in SIR-epidemics.} Scaling of the infected mass $I$ vs $r$, calculated as discussed in Ref.~\cite{Li2011}. As expected, $I$ undergoes the same geometric crossover and critical stretching of $M$ as in Eq.~\eqref{eq:4}. (Inset) Spreading velocity of the epidemic front wave, scaled according to Eq.~\eqref{eq:5}. (Sub-Inset) Linear fit of $\tau_*=\zeta^{\varpi}$ yielding $\varpi\approx0.99$. 
	Results are for $N=2.5\times10^7$, and averaged over $10^4$ realizations.}
	\label{fig:4}\vspace{-0.5cm}
\end{figure}
On the contrary, $\ell$ and $\langle r\rangle$ share the same metric in the lattice regime, and so $\langle r\rangle\sim\ell$ for $\langle r\rangle\gtrsim \xi_\times$. 
One can then advance the scaling $\langle r\rangle= \zeta\ell^{1/2}\mathpzc{f}(\ell)$ where $\mathpzc{f}(\ell)\propto\ell^{1/2}$ for $\ell\gtrsim\xi_\ell^\times(\zeta)$ and constant otherwise.
The quantity $\xi_\ell^\times$ can be estimated as the chemical length up to which the small-world extends, so that $\xi_\ell^\times\!\propto\!\log N_\zeta\!\sim\!\log\zeta$ as $N_\zeta\!\sim\!\zeta^2$. 
The data collapse in Fig.~\ref{fig:3}(d) is consistent with our scaling form, and supports (Fig.~\ref{fig:3}(d), inset) the approximation $\xi_\ell^\times\!\sim\log\zeta$. 
Altogether, these results confirm the intuition that far from criticality the network's geometry is governed by the characteristic scale $\zeta$, corroborating the surprising effect of the critical stretching.

\paragraph*{\underline{Stretching effects in spreading dynamics}.} 
$\!\!\!\!$Since critical phenomena are strongly affected by the geometry of the underlying interaction structure, the critical stretching will accordingly influence the dynamics of processes acting on the network.
As an illustrative case, let us consider an SIR process with infection rate $\beta$ and recovery time $\gamma\equiv1$. 
In light of the mapping to bond percolation~\cite{epidemics}, the clusters of nodes removed by the outbreak are geometrically equivalent to those originated from a static deletion, whereas now the shortest-path $\ell$ identifies the number of time steps $t$ separating infected nodes in the $\ell^\mathrm{th}$ shell from an initial spreader. 
Setting the relative rate $\beta'\equiv1-\beta/\beta_c$ to a value $\beta'\ll\zeta^{-1}$, we thus expect the spreading of the disease to undergo a {\em temporal crossover} at the intrinsic time scale $\tau_*\equiv\lfloor\zeta\rfloor$ from a mean-field behavior to a $2D$ one. 
In particular, interpreting $\langle r\rangle$ as the average distance of the infective nodes from an initial spreader, we can deduce from Eq.~\eqref{eq:3} that the spreading velocity $\mathpzc{v}:=\mathrm{d}\langle r\rangle/\mathrm{d} t\sim t^{1/d_{min}-1}$ of the outbreak front will accordingly satisfy the temporal scaling
\begin{equation}\label{eq:5}
\frac{\mathpzc{v}\big(t,\zeta\big)}{\mathpzc{v}_*(\zeta)}=\bigg(\frac{t}{\tau_*(\zeta)}\bigg)^{\!\!-\nicefrac{1}{2}}
\mathcal{V}\bigg(\frac{t}{\tau_*(\zeta)}\bigg),
\end{equation}
\noindent
where $\mathcal{V}(\mathpzc{x})\propto\mathpzc{x}^{1/d_{min}^{2D}-1/2}$ for $\mathpzc{x}\gg1$ and constant otherwise, and $\mathpzc{v}_*(\zeta)\equiv\lfloor\zeta\rfloor^{1/2}$ is the crossover velocity. 
The data collapse shown in Fig.~\ref{fig:4} validates our premises, confirming that, if critical, the epidemic front undergoes a spatiotemporal crossover at the universal distances $(\xi_*,\tau_*)$ from the infective spreader, varying from a diffusive to a super-diffusive propagation.
Considering that the crossover time scale changes from a logarithmic to a linear dependence with $\zeta$---since $\tau_\times\!\!\sim\!\log\zeta$ far from criticality---the critical stretching results in an intriguing scenario: if critical, an outbreak will invade the system at ``slow'' (i.e.\,diffusive) rates for longer times, before turning super-diffusive. 
Besides being of theoretical interest, these results are relevant for studying spreading processes in real-world systems, e.g. in virus propagation on wireless networks~\cite{Wire009} or on the Internet~\cite{Zeg996,Watts2002}, or in the spreading of diseases through transportation networks~\cite{Hal014,Dan016} and of rumors in social networks~\cite{Watts2002}, whose spatial constraints are well captured by the exponential wiring cost function here adopted. 

\paragraph*{\underline{Outlook.}\!\!\!} 
We have investigated the effects that a tunable characteristic link length $\zeta$ has on the universality of critical phenomena in spatial networks.
As a theoretical benchmark, we adopted a structurally disordered lattice characterized by an exponential distribution of link-lengths, although our results are expected to hold generally for systems with a finite range of interactions and homogeneous distributions of connectivity. 
We found that near the percolation threshold, the random structure (otherwise bounded by the characteristic scale $\zeta$) expands until the intrinsic length scale $\xi_*=\zeta^{3/2}$ before crossing over to the spatial one, a phenomenon we named {\em critical stretching}. 
Besides affecting the geometry of the incipient cluster, the critical stretching prolongates the duration of the mean-field regimes featured by dynamical processes occurring on it, as we have demonstrated on SIR epidemics.
Since these spatiotemporal effects become stronger at higher (embedding) dimensions---e.g., in SIR $\tau_d^*(\zeta)=\lfloor\zeta\rfloor^{2d/(6-d)}$ and $\xi_d^*(\zeta)=\zeta^{6/(6-d)}$ for $2\leq d<6$---the critical stretching should be experimentally verifiable. 
In fact, besides accurately modeling transport~\cite{Hal014,Dan016} and communication networks~\cite{Wire009,Zeg996,Yook2002,Watts2002}, recent evidences~\cite{Brain1,Brain2,Brain3} support the existence of scale-invariant organization principles~\cite{Brain3} in mammal brains based on the exponential wiring cost function examined here.
In this respect, our results offer an important keystone for understanding the neural functioning---e.g.\,in reference to the observed propagation of activity before epileptic seizures~\cite{Vlad011}---and they raise the tantalizing opportunity of developing some criteria (e.g.\,by measuring the amount of stretching in the spatio-temporal propagation of the action potentials) for testing the ``critical brain hypothesis''~\cite{Mor013}.

\paragraph*{\underline{Acknowledgments.}\!\!\!} I. B. and B. G. contributed equally to this work. 
S.H.\,acknowledges financial support from the ISF, ONR Grant, N62909-14-1-N019; DTRA Grant: HDTRA-1-10-1-0014, BSF-NSF No.\,2015781, MOST with JSF, MOST with MAECI, the Army Research Office, the BIU Center for Research in Applied Cryptography and Cyber Security. 
I. B. thanks S. V. Buldyrev, G. Sicuro, M. C. Strinati for valuable discussions. 


\begin{thebibliography}{99}
\bibitem{note1} The upper bound $\zeta_{max}=L/2$ reflects the choice of periodic boundary conditions, while the lower one $\zeta_{min}=1$ describes the lattice dominated regime, since $\rho\sim1/a^2$ for $L\gg1$, where $a\equiv1$ is the lattice spacing.
\bibitem{note2} In purely random structures (e.g.\,Bethe lattice), $\xi$ is itself dimensionless being defined as a formal embedding (see e.g.\,Ref.~\cite{BuH012}, $\mathsection2.4.2$) of $\xi_\ell$ (where $\ell$ is the minimal number\vspace*{-0.06cm} 
\noindent 
of hops) onto a $d\geq6$ Euclidean space s.t.\,$\xi_{MF}\sim\xi_\ell^{1/2}$.
\end{thebibliography}
\end{document}